\documentclass[a4paper,fleqn,usenatbib,useAMS]{mnras}

\usepackage{graphicx}	
\usepackage{amsmath}	
\usepackage{amssymb}	
\usepackage{multicol}        
\usepackage{bm}		
\usepackage{pdflscape}	

\newcommand{\kms}{\,km\,s$^{-1}$} 

\usepackage{gensymb} 
\usepackage{soul} 

\def\kms{\,{\rm km~s^{-1}}}

\def\ltsima{$\; \buildrel < \over \sim \;$}
\def\simlt{\lower.5ex\hbox{\ltsima}}
\def\gtsima{$\; \buildrel > \over \sim \;$}
\def\simgt{\lower.5ex\hbox{\gtsima}}

\def\ltsima{$\; \buildrel < \over \sim \;$}
\def\simlt{\lower.5ex\hbox{\ltsima}}
\def\gtsima{$\; \buildrel > \over \sim \;$}
\def\simgt{\lower.5ex\hbox{\gtsima}}

\title[Kinematic trends in young and old stars]{Correlations between age, kinematics, and chemistry as seen by the RAVE survey}

\author[J.~Wojno et al.]{
Jennifer Wojno,$^{1,2}$\thanks{E-mail: jwojno1@jhu.edu (JW)} 
Georges Kordopatis,$^{3}$
Matthias Steinmetz,$^{1}$
Paul McMillan,$^{4}$
\newauthor
James Binney,$^{5}$
Benoit Famaey,$^{6}$
Giacomo Monari,$^{1}$
Ivan Minchev,$^{1}$
\newauthor
Rosemary F.\ G.\ Wyse,$^{2}$
Teresa Antoja,$^{7}$
Arnaud Siebert,$^{6}$
Ismael Carrillo,$^{1}$
\newauthor
Joss Bland-Hawthorn,$^{8}$
Eva K.\ Grebel,$^{9}$
Toma\v z Zwitter,$^{10}$
Olivier Bienaym\'e,$^{6}$
\newauthor
Brad Gibson,$^{11}$
Andrea Kunder,$^{12}$
Ulisse Munari,$^{13}$
Julio Navarro,$^{14}$
Quentin Parker,$^{15}$
\newauthor
Warren Reid,$^{16,17}$
George Seabroke$^{18}$
\\
$^{1}$Leibniz Institut f\"ur Astrophysik Potsdam, An der Sternwarte 16, 14482 Potsdam, Germany\\
$^{2}$Department of Physics and Astronomy, Johns Hopkins University, 3400 N. Charles St, Baltimore, MD 21218, USA \\
$^{3}$Laboratoire Lagrange, Universit\'e C\^ote d'Azur, Observatoire de la C\^ote d'Azur, Boulevard de l'Observatoire, CS 34229, 06304, Nice, France\\
$^{4}$Lund Observatory, Lund University, Department of Astronomy and Theoretical Physics, Box
43, SE-22100 Lund, Sweden \\
$^{5}$Rudolf Peierls Centre for Theoretical Physics, Keble Road, Oxford OX1 3NP, UK\\
$^{6}$Observatoire astronomique de Strasbourg, Universit\'e de Strasbourg, CNRS, UMR 7550, 11 rue de l'Universit\'e, F-67000 Strasbourg, France \\
$^{7}$Dept. FQA, Institut de Ciencies del Cosmos (ICCUB), Universitat de Barcelona (IEEC-UB), Marti Franques 1, 08028 Barcelona, Spain \\
$^{8}$Sydney Institute for Astronomy, School of Physics A28, University of Sydney, NSW 2006, Australia\\
$^{9}$Astronomisches Rechen-Institut, Zentrum f\"ur Astronomie der Universit\"at Heidelberg, M\"onchhofstr. 12-14, D-69120 Heidelberg, Germany\\
$^{10}$Faculty of Mathematics and Physics, University of Ljubljana, 1000 Ljubljana, Slovenia\\
$^{11}$E.A. Milne Centre for Astrophysics, University of Hull, Hull HU6 7RX, UK\\
$^{12}$Saint Martin's University, 5000 Abbey Way SE, Lacey, WA 98503, USA\\
$^{13}$INAF National Institute of Astrophysics, Astronomical Observatory of Padova,  36012 Asiago (VI), Italy\\
$^{14}$Senior CIfAR Fellow, Department of Physics and Astronomy, University of Victoria, Victoria, BC, Canada V8P 5C2\\
$^{15}$Department of Physics, Chong Yuet Ming Physics Building, The University of Hong Kong, Hong Kong\\
$^{16}$Department of Physics and Astronomy, Macquarie University, Sydney, NSW 2109, Australia\\
$^{17}$Western Sydney University, Locked Bag 1797, Penrith South, NSW 2751, Australia\\
$^{18}$Mullard Space Science Laboratory, University College London, Holmbury St Mary, Dorking, RH5 6NT, UK
}

\date{Accepted XXX. Received YYY; in original form ZZZ}

\pubyear{2018}

\begin{document}
\label{firstpage}
\pagerange{\pageref{firstpage}--\pageref{lastpage}}
\maketitle

\begin{abstract}

We explore the connections between stellar age, chemistry, and kinematics across a Galactocentric distance of $7.5 < R\,(\mathrm{kpc}) < 9.0$, using a sample of $\sim 12\,000$ intermediate-mass (FGK) turnoff stars observed with the RAdial Velocity 
Experiment (RAVE) survey. The kinematics of this sample are determined using radial velocity measurements from 
RAVE, and parallax and proper motion measurements from the Tycho-Gaia Astrometric Solution (TGAS). In addition, 
ages for RAVE stars are determined using a Bayesian method, taking TGAS parallaxes as a prior. 
We divide our sample into young ($0 < \tau < 3$ Gyr) and old 
($8 < \tau < 13$ Gyr) populations, and then consider different metallicity bins for each of these age 
groups. We find significant differences in kinematic trends of young and old, metal-poor and metal-rich, stellar 
populations. In particular, we find a strong metallicity dependence in the mean Galactocentric radial 
velocity as a function of radius ($\partial {V_{\rm R}}/\partial R$) for young stars, with metal-rich stars having a much 
steeper gradient than metal-poor stars. For $\partial {V_{\phi}}/\partial R$, young, metal-rich stars significantly lag the 
LSR with a slightly positive gradient, while metal-poor stars show a negative gradient above the LSR. We interpret these 
findings as correlations between metallicity and the relative contributions of the non-axisymmetries in the Galactic gravitational 
potential (the spiral arms and the bar) to perturb stellar orbits.
\end{abstract}





\section{Introduction}
\label{sec:introduction}

The field of Galactic archaeology stands poised to reveal the formation
history of our Galaxy. As low- and intermediate-mass stars are long-lived,
they can act as time capsules, allowing us a window to the environment in
which they were born \citep{Freeman02}. Spectroscopic surveys, such as RAVE
\citep{Steinmetz06}, SEGUE \citep{Yanny09}, APOGEE \citep{Majewski17},
Gaia-ESO \citep{Gilmore12}, LAMOST \citep{Zhao12}, and GALAH
\citep{DeSilva15}, provide a number of crucial parameters necessary for
disentangling the formation history of the Galactic disc, such as stellar
radial velocities, effective temperatures, surface gravities, and individual
chemical abundances. The precision with which we can exploit these spectra is
being greatly increased by the progressive release of astrometric data from
ESA's Gaia satellite, starting with Gaia DR1 in September 2016 \citep{Gaia16,
Lindegren16} and continuing with Gaia DR2 in April 2018.

It has been known for more than a
century that the local velocity field is not uniform but contains moving
groups \citep[e.g.][]{Proctor69,Kapteyn05}\footnote{For a history of the
discovery of moving groups see \citet{Antoja10}.}: predominantly the Hyades,
Pleiades, and Hercules groups. The precision of stellar velocity measurements in the solar neighbourhood 
has since increased dramatically, 
and as a result it is now possible to study this velocity space in fine detail 
 \citep[e.g.,][]{Dehnen00,Famaey08,Williams13,Antoja15,Antoja17,Kushniruk17,Carrillo18}. 
These structures are indicative of deviations of the Milky Way (MW) disc from
an idealized axisymmetric model associated with our Galaxy's bar and
spiral arms. The way such structures vary in velocity space as a function  
Galactocentric radius has been explored in a number of simulations
\citep[e.g.,][]{Minchev10,Quillen11,Antoja11,McMillan13,Monari14,Monari17_df},
with a view to constraining the nature of the underlying non-axisymmetries. 
To understand better the dynamical processes that shape the kinematic trends
we see in stellar populations today, we turn to measurements of the local
mean stellar velocity field $\mathbf{V}$, which we analyse in Galactocentric
cylindrical polar coordinates $(R,\phi,z)$.

\citet{Siebert11} first detected a gradient in $V_R$, namely $\partial
{V_{\rm R}}/\partial R\lesssim-3$\,km\,s$^{-1}$\,kpc$^{-1}$, using
the line-of-sight velocities of RAVE stars. \citet{Siebert12} fitted this
shallow gradient to a model on the assumption that it is solely due to
long-lived spiral arms. They found that they were able to reproduce the
observational results with a two-armed model in the solar neighbourhood ($d <
1\,$kpc), although they acknowledged that the bar could contribute.
\citet{Monari14} developed a model of the bar's contribution to the velocity
field near the Sun. They showed that if the Sun is located close to the bar's
outer Lindblad resonance (OLR), from our position we would measure a trend
similar to that found in \citet{Williams13}, which measured a slightly
steeper gradient, $\partial V_{R}~/~\partial
R=-8$\;km\,s$^{-1}$\,kpc$^{-1}$. \citet{Grand12} analysing N-body simulations
and \citet{Monari16}, modelling spiral arms with linear perturbation theory,
showed that stars located in arms move in toward the Galactic centre, while
stars in interarm regions move outwards. \citet{Monari16} estimated the gradient
in the data to be $\partial V_R/\partial
R\simeq-8$\,km\,s$^{-1}$\,kpc$^{-1}$.

The study by \citet{Faure14} of the combined effect of the bar and spiral
arms noted that the pattern just described in which stars in arms move
inwards reverses outside the corotation resonance, so there stars in arms
move outwards while stars between arms move inwards. This finding is
consistent with the perturbative model of \citet{Monari16}.

Soon after the gradient in $V_R$ was found in RAVE stars, \citet{Widrow12}
discovered significant deviations in $V_z$ with height $z$ above the plane
using the Sloan Digital Sky Survey (SDSS). Using the high-resolution
simulation of \citet{Purcell11} to study of the effect of the Sagittarius
dwarf spheroidal galaxy (dSph) on the MW disc, \citet{Gomez13} concluded that by plunging
through the disc $\sim2\,$Gyr ago, the Sagittarius dSph could have generated the
pattern in $V_z$ detected by \citet{Widrow12}.  \citet{Carlin13} found a
similar signature in LAMOST stars that are distributed through a larger
volume.  \citet{Williams13} analysed the $V_z$ field defined by RAVE stars,
and detected wave-like motions perpendicular to the plane that could either
be associated with spiral arms or with a recent accretion event. However,
\cite{Carrillo18}, reanalysing the RAVE data, laid bare the extent to which
the small measured values of $V_z$ are vulnerable to systematic errors in
both proper motions and distances. Using TGAS proper motions and improved
distances, they concluded that vertical motions inside the solar
radius could be induced by the Galactic bar or spiral arms, while outside
they are likely generated by a recent satellite/merging event.

Recently, \citet{Antoja17} investigated the dependence of the velocity field
on metallicity.  In a novel analysis applied to data from RAVE and the
Geneva-Copenhagen Survey \citep[GCS,][]{Holmberg07} they compared the
metallicity distribution at ($V_{R},V_{\phi}$) with that at ($-V_{
R},V_{\phi}$), which would be identical in a well-mixed axisymmetric model. 
The fact that the observed metallicity distributions differed is consistent
with previous studies \citep[e.g.,][]{Famaey07} that have shown that the
chemical abundances of moving groups differs from that of the local interstellar medium
(ISM). In particular, moving groups are not chemically homogeneous, so they
cannot be dissolved open clusters.

A significant barrier to uncovering the chemodynamical history of the solar
neighbourhood is the difficulty of determining the ages of individual field stars. The
classical approach to age determination involves measuring the luminosity of
turnoff stars of known colour and metallicity. This method requires an
accurate distance. We now have trigonometric parallaxes from the Tycho-Gaia
Astrometric Solution (TGAS) \citep{Gaia16, Lindegren16} for two million
stars, so we can determine credible ages for large samples of stars. Here we
combine TGAS parallaxes and proper motions with RAVE spectroscopy of turnoff stars to discover
how the velocity field depends on age and metallicity.

In Section~\ref{sec:data} we review the RAVE survey and its
updated distance pipeline from which our distances and ages are derived
(Section~\ref{sec:updated_distances}). In Section~\ref{sec:age_validation} we
record the criteria used to select our samples and the steps taken to
validate our ages.  With our sample of stars in hand, we then explore
kinematic trends, in particular $\partial {V_{R ,\phi}}/\partial R$
for two selected age bins (young and old) in Section~\ref{sec:velocity_dist}, as
a function of Galactocentric radius. Section~\ref{sec:orbital_parameters} 
presents the distributions of orbital parameters for our age-metallicity bins. 
In Section~\ref{sec:discussion}
we present a discussion and interpretation of our results, and draw our
conclusions from this analysis.

\section{The data}
\label{sec:data}
\subsection{The RAVE survey} \label{sec:RAVE}

The RAVE survey collected over half a million medium-resolution ($R \approx 7500$) spectra of stars in 
the Southern hemisphere from 2003 until 
2013, using the 6dF multi-object spectrograph on the 1.2-m UK Schmidt Telescope at the Siding Spring 
Observatory in Australia. Throughout the course of the survey, parameters derived from the spectra were made publicly available 
via a number of data releases, with the fifth data release (DR5) being the latest \citep{Kunder17}, 
providing 520\,781 measurements for 457\,588 individual stars. Centred on the Ca\,\textsc{ii}-triplet 
region (8410~--~8795~\AA) region, this spectral range was chosen specifically to coincide with the spectral 
range of Gaia's Radial Velocity Spectrometer (RVS)  \citep{Prusti12,Bailer-Jones13,Recio-Blanco16}. 

In addition to radial velocities \citep[typical uncertainties $\sim2 \kms$,][]{Kunder17,Kordopatis13}, RAVE DR5 contains a number 
of other stellar parameters derived from spectra including effective temperature, surface gravity, metallicity ([M/H]), as well as 
individual abundances for six elements (Mg, Al, Si, Ti, Fe, Ni) \citep{Boeche11}. To provide additional parameters 
such as apparent magnitudes and proper motions, RAVE DR5 was 
cross-matched with a number of other astrometric and photometric catalogues. In 
particular, RAVE has a significant overlap with stars available in TGAS \citep{Lindegren16}, with RAVE DR5 containing 215\,590 unique TGAS stars. Currently, the RAVE survey offers one of the largest 
samples of stars with accurate 6D phase-space information in addition to stellar parameters derived from spectra.

\begin{figure*}
 \includegraphics[width=0.9\textwidth]{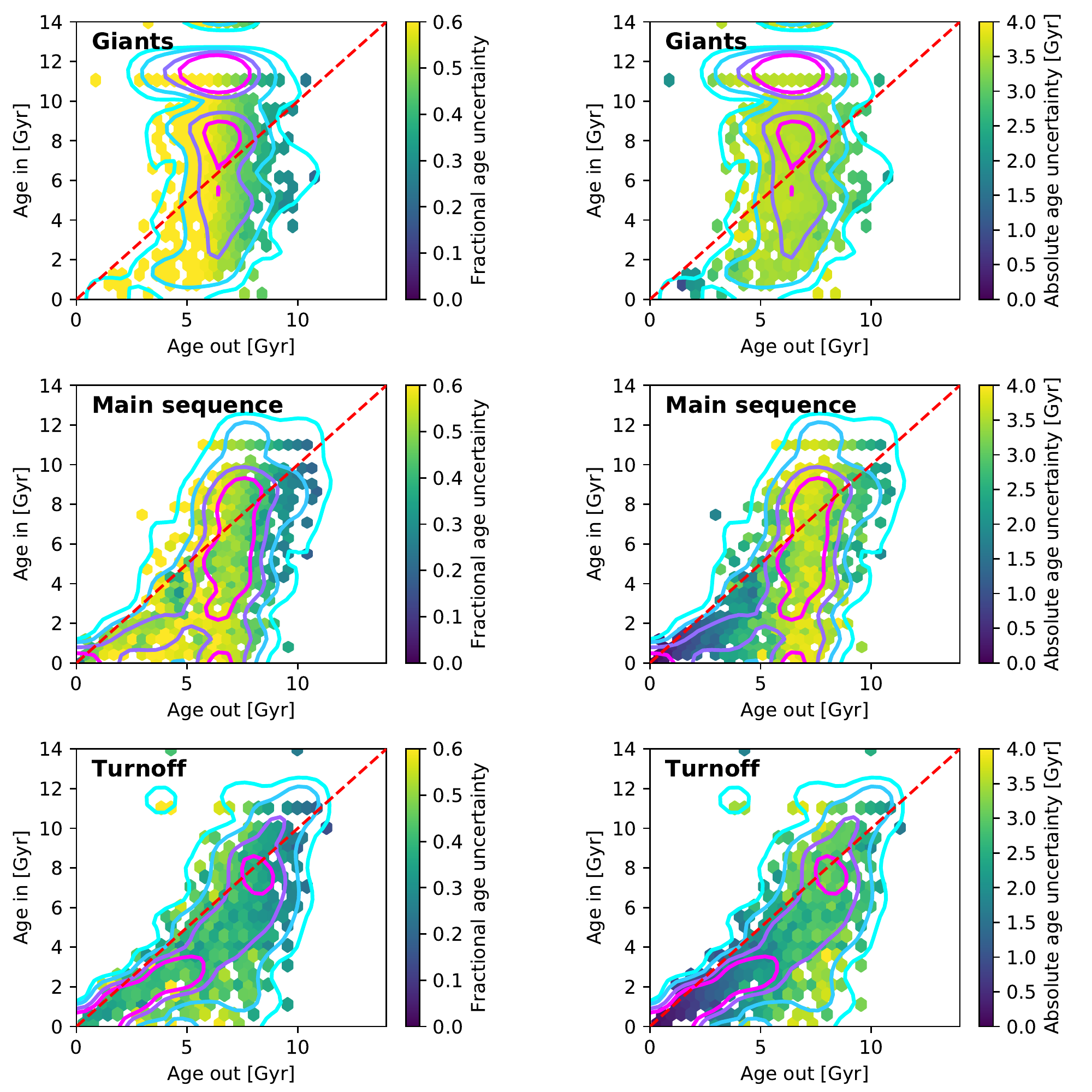}
 \caption{Input age vs. output age, with bins coloured by fractional age uncertainty in the left column, and by absolute age uncertainty in the right column. Giants ($T_{\rm eff} < 5500$\,K, log $g < 3.5$), main sequence stars (log $g > 4.25$), and turn off stars $(T_{\rm eff} > 5500$\,K, $3.5 < \mathrm{log}\,g < 4.25)$ are shown as the top, middle, and bottom rows, respectively. Contours indicate 33, 67, 90, and 99 per cent of the sample.}
  \label{fig:age_comparison}
\end{figure*}

\subsection{Distances and ages}
\label{sec:updated_distances}

We use distances and ages returned by an updated version of the pipeline
described in \citet{Binney14}, details of which are given in
\citet{McMillan17}. The updated pipeline allows for TGAS parallaxes to be
included as priors, alongside stellar parameters such as temperature, surface
gravity, and metallicity from RAVE, apparent magnitudes from 2MASS
\citep{Skrutskie06} and AllWISE \citep{Cutri13}, and an underlying Galactic
model. The default Galactic model (prior) used is the same as that used for
the distance determinations available in DR4 and DR5 and includes priors on
stellar metallicities and ages for a given disc component \citep[][Section 2.1]{McMillan17}. Details of the 
other available priors can be found in Section 5 
of \citet{McMillan17}. From both internal tests and comparisons with external
catalogues, \citet{McMillan17} report that the combined spectrophotometric
distance estimates perform better than purely spectroscopic or astrometric
estimates.

In addition to distance estimates, the updated pipeline produces
estimates of stellar mass, metallicity, line of sight extinction, and age as
byproducts. For this work, we want to avoid a prior where a relationship between age
and metallicity is imposed. Therefore, we use age estimates derived 
utilizing a Galactic model with a flat prior on both age and metallicity \citep[`Density' in][]{McMillan17}.
The ages derived using this prior are roughly consistent with those derived using an age prior where the 
star formation rate declines over time \citep[][Equation 20]{McMillan17}, and a flat prior in metallicity. 
Derived distances are not significantly affected by the choice of prior.

\subsection{Age validation}
\label{sec:age_validation}

To validate the pipeline's age estimates, we generate a mock catalogue of
RAVE-like stars using the population synthesis code \textsc{Galaxia}
\citep{Sharma11}. As \textsc{Galaxia} uses PARSEC \citep{Bressan12}
isochrones to produce stars with perfectly-known ages and atmospheric
parameters ($T_{\rm eff}$, $\log g$, [M/H]), 2MASS photometry, and distances,
it offers a suitable test sample for assessing internal
uncertainties on the output of the distance pipeline. 

\subsubsection{Mock catalogue generation}

We generated stars with I-magnitude
range $7 < \mathrm{I}_{\mathrm{DENIS}} < 13$, to cover the whole magnitude
range of RAVE, and removed all stars with Galactic latitudes $|b| <
5^{\circ}$. We then resampled the age distribution of our mock catalogue
(primarily by reducing the number of young stars) to ensure we had enough
stars for robust statistics across the entire range of stellar ages, and as a
side-effect, roughly reproduce the age distribution of the extended solar
neighbourhood.

\subsubsection{Applying uncertainties to our mock catalogue}

We then scattered the mock data for $T_{\rm eff}$, log \,$g$, and [M/H] by
RAVE-like uncertainties. The standard deviations used in this process were
the quadrature sums of RAVE's internal and external uncertainties for a star
with the given true data \citep[for more detail see Figure 5 of][]{Wojno17}.

The true apparent magnitudes $J$, $H$ and $K_s$ in the mock data were then
scattered by a Gaussian error distribution with dispersion $0.025$ mag, which is
a typical 2MASS uncertainty. Finally, the true parallaxes were scattered by
a Gaussian error distribution with a dispersion of $0.3\,$mas, which is a
typical TGAS uncertainty \citep{Gaia16,Lindegren16}. 

\subsubsection{Input versus output age comparison}
\label{sec:comparison_age_bins}

The mock data were then fed into the updated RAVE distance pipeline and the resulting
ages compared with the `true' ages given by \textsc{Galaxia}.  As a
preliminary sanity check the values of $T_{\rm eff}$, $\log g$, [M/H] and
distance from the pipeline were compared with the original values from
\textsc{Galaxia} and found to agree within the errors.
Figure~\ref{fig:age_comparison} shows the input and output ages subdivided by
regions within the $T_{\rm eff}-\log g$ plane. The top row shows the data for
giant stars ($T_{\rm eff} < 5500$\,K, $\log g < 3.5$), the middle row shows
the data for main-sequence stars ($\log g > 4.25$), and the bottom row shows
the data for turnoff stars $(5500\,{\rm K} < T_{\rm eff} < 7100$\,K, $3.5 <
\log g < 4.25)$. Panels in the left column are colour-coded by fractional age
uncertainty, while panels on the right are colour-coded by absolute age
uncertainty. The contours enclose 33, 67, 90, and 99 per cent of the sample.
Unsurprisingly, the giant and main-sequence samples have much higher
uncertainties than the turnoff sample, with almost all stars being assigned
an intermediate age regardless of its true age. Consequently, we exclude all
but turnoff stars from further consideration.  For turnoff stars, the
isochrones are much better separated than for the giant and lower
main-sequence stars, and therefore yield smaller age uncertainties (see
Figure~\ref{fig:teff_logg_age_err}).

In order to obtain a balanced sample with the most precise age estimates
possible, we restrict further analysis to turnoff stars in two age bins:
young ($0 < \tau < 3\,$Gyr) and old ($8 < \tau < 13\,$Gyr) stars. We do not
use stars of intermediate-age ($3 < \tau < 8\,$Gyr) because this age group
manifests a significant systematic offset. While this offset disappears if we
only consider stars with age uncertainties less than 20\%, we find
that this prunes the sample in a biased way. Intermediate-age stars are more
likely to have large age uncertainties because their
errors are unrestricted: a young star cannot have an age smaller than zero,
and an old star cannot have an age that exceeds $13.8\,$Gyr. 

\subsubsection{Contamination}

\begin{figure}
\centering
 \includegraphics[width=0.8\columnwidth]{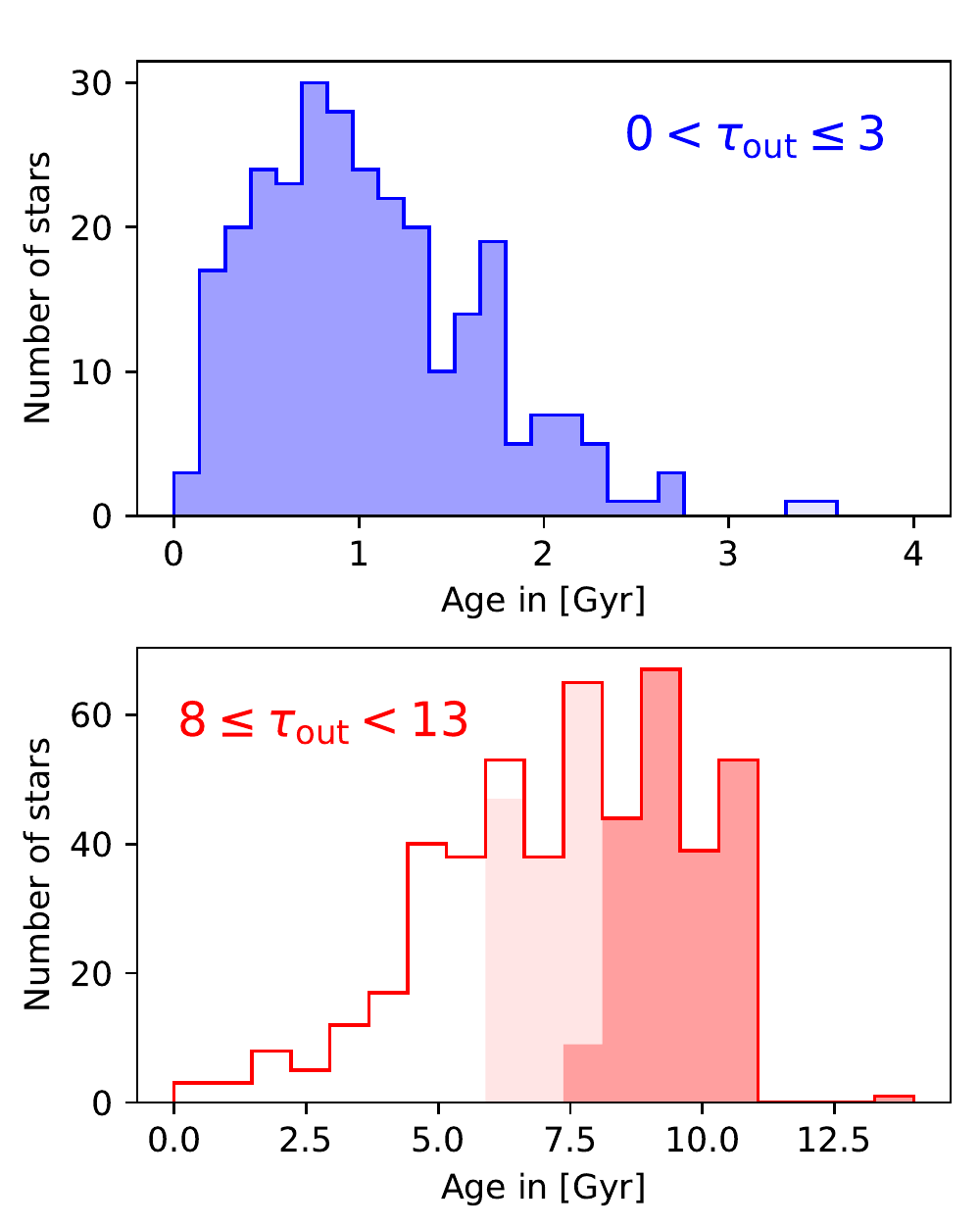}
 \caption{Distribution of input ages for the stars with output ages given by the range indicated in the plot. The top panel shows the distribution for our `young' sample, while the bottom panel shows the distribution for our `old' sample. The darker shaded regions indicate stars with input ages that fall within the given bin, while the lighter shaded regions indicate stars with input ages that fall slightly outside of the given bin. A lack of shading indicates stars which are considered contaminants of that bin.}
  \label{fig:age_bin_distributions}
\end{figure}

\begin{figure}
\centering
 \includegraphics[width=\columnwidth]{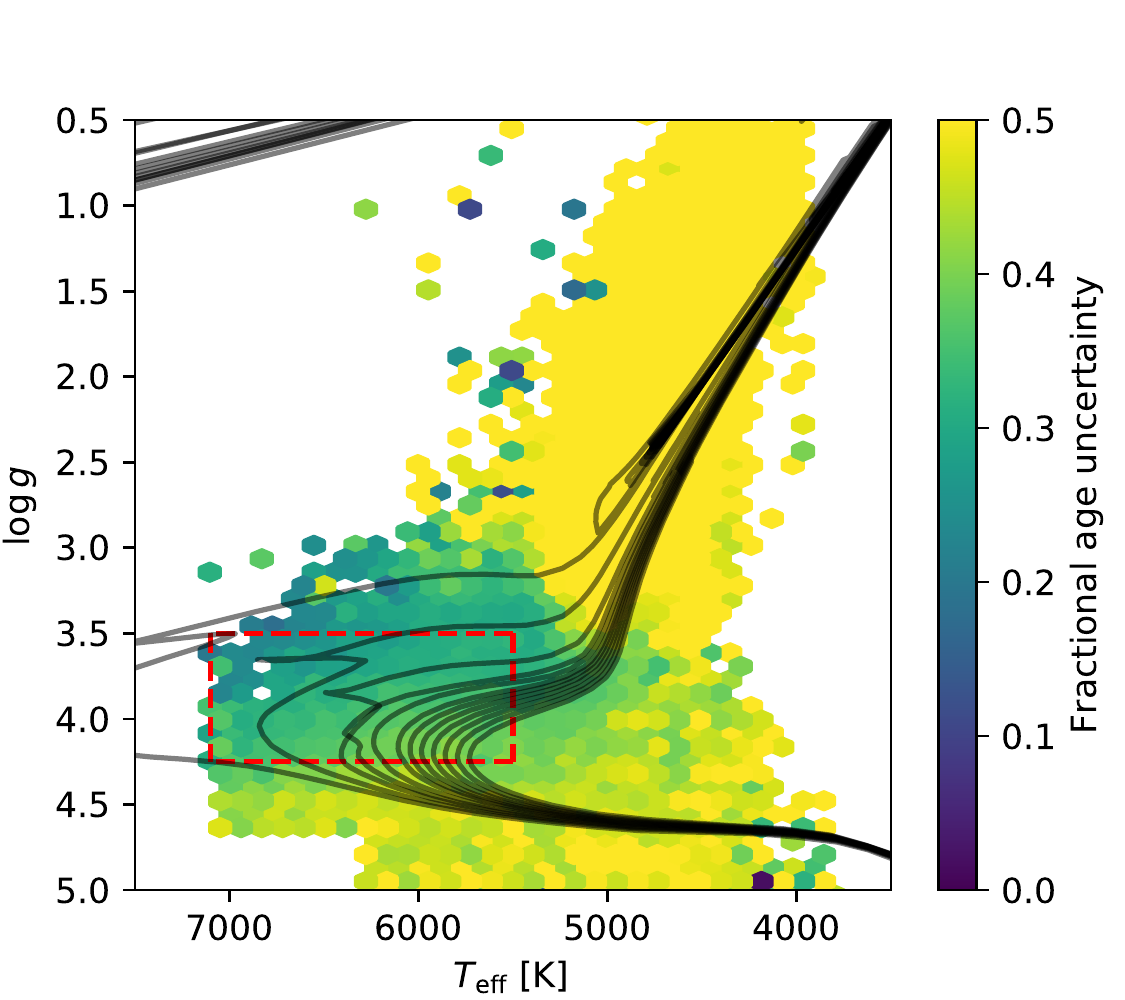}
 \caption{Spectroscopic $T_{\rm eff}-$log\,$g$ diagram of RAVE DR5 stars which satisfy the quality criteria listed in Sec.~\ref{sec:RAVE_qc}. The dashed red lines indicate the cuts made in the parameter space to select for only turnoff stars. The bins are colour-coded by the fractional age uncertainty. Solar-metallicity isochrones are overplotted in black, and span a range in age from 1 to 13 Gyr, with a step size of $\sim$1 Gyr.}
 \label{fig:teff_logg_age_err}
\end{figure}

\begin{figure}
\centering
 \includegraphics[width=0.8\columnwidth]{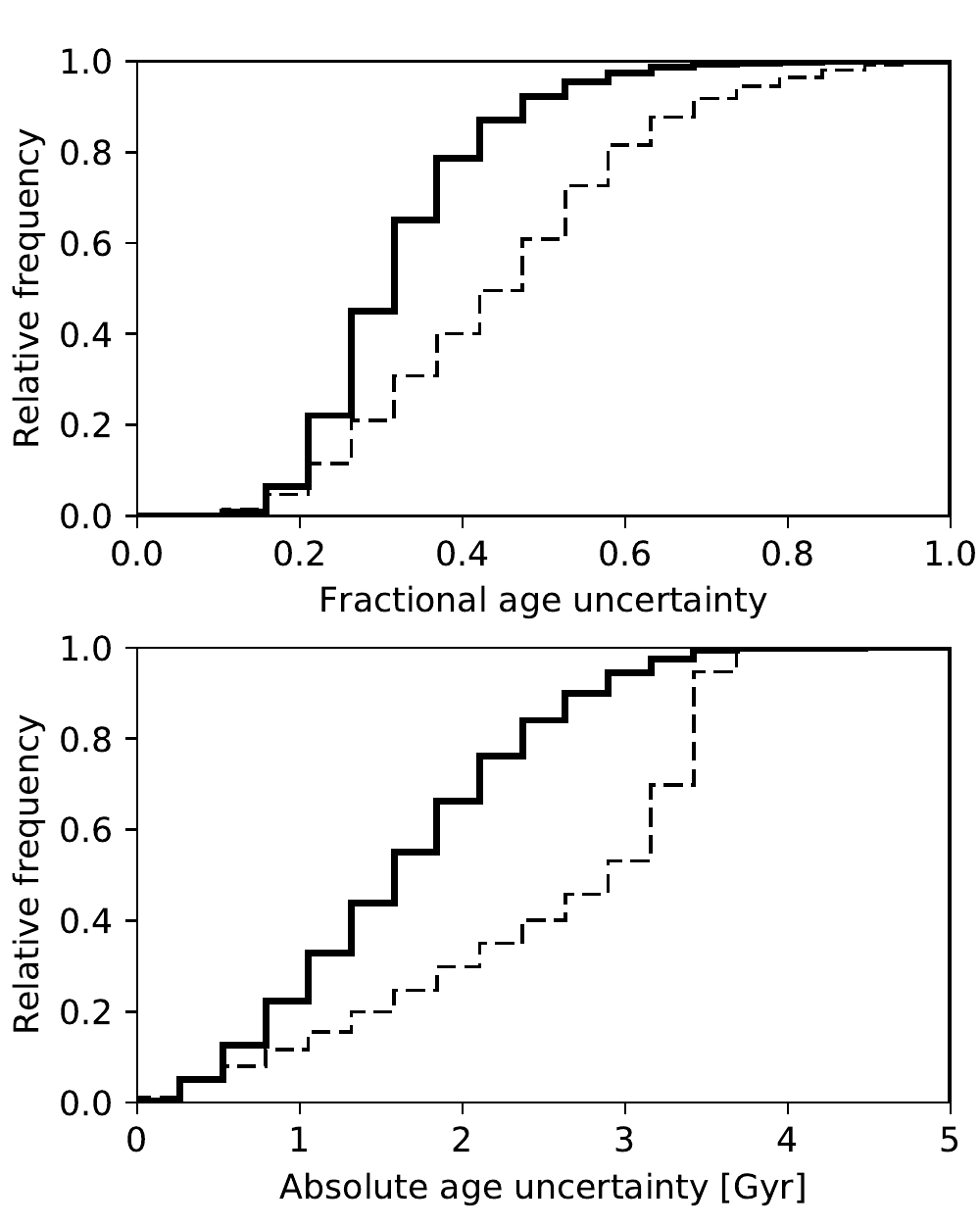}
 \caption{Cumulative histogram of the fractional (top) and absolute (bottom) age uncertainties of our final selected sample of turnoff stars (solid line), and for the whole RAVE-TGAS sample (dashed line).}
 \label{fig:age_err_hist}
\end{figure}

We now estimate the number of stars that wrongly appear in the `young' and
`old' bins as a consequence of observational and theoretical uncertainties.
In Figure~\ref{fig:age_bin_distributions},
we show the distributions of input ages of stars that end up in the young ($0
< \tau < 3\,$Gyr) and old ($8 < \tau < 13\,$Gyr) bins. The dark-shaded areas
are regions within which the star has been correctly classified as young or
old. The lighter-shaded areas indicate regions in which the star is not as
young or old as its bin membership implies, but is nonetheless pretty young
($\tau \leq 4\,$Gyr) if it is in the young bin, or pretty old ($\tau \geq
6\,$Gyr) if it is in the old bin.  From this analysis we conclude that our
`young' sample has a success rate of $\sim 98$ per cent in the sense that
less than $\sim 2$ per cent of its stars are older than $4\,$Gyr, and our
`old' sample has a success rate of $\sim 73$ per cent in the sense that less
than $\sim 27$ per cent of its stars are younger than $6\,$Gyr.

\subsection{Selection of our RAVE-TGAS sample}

In light of the work presented in the last section, henceforth we confine our 
analysis to the 37\,765 RAVE turnoff stars ($5500 < T_{\rm eff} < 7100$\,K, $3.5 < \log
g < 4.25$). This selection in $T_{\rm eff}-$log$\,g$ space is shown in
Figure~\ref{fig:teff_logg_age_err} by the dashed red lines. Pixels in
Figure~\ref{fig:teff_logg_age_err} are colour-coded by fractional age
uncertainty, with solar-metallicity isochrones overplotted in black, spanning
a range in age from 1 to $13\,$Gyr in steps of 1 Gyr. 

A histogram of the age uncertainties for this sample is shown in
Figure~\ref{fig:age_err_hist}. The majority of our sample ($\sim 60$ per
cent) has age uncertainties less than $2\,$Gyr, with a sizable fraction ($\sim
25$ per cent) having age uncertainties less than $1\,$Gyr.

\subsubsection{RAVE quality criteria} 
\label{sec:RAVE_qc}

\begin{figure}
\centering
 \includegraphics[width=\columnwidth]{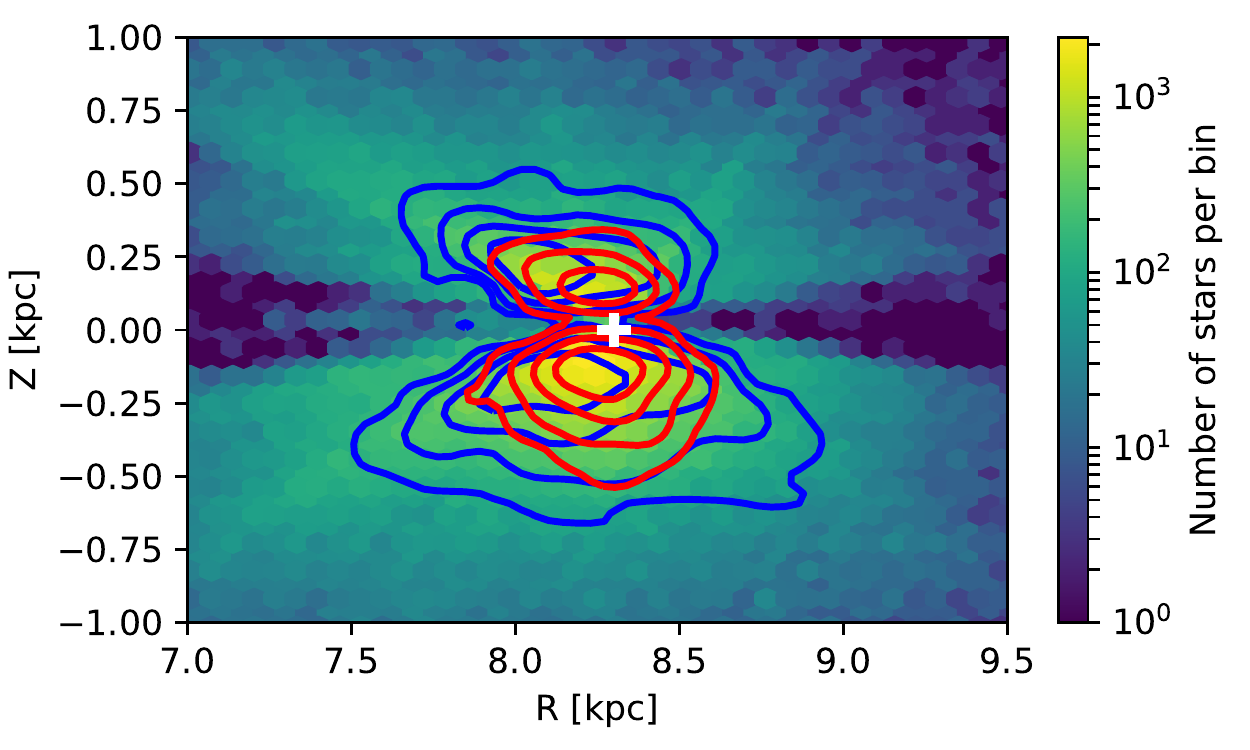}
 \caption{Spatial extent of the selected young (blue contours) and old (red contours) turnoff stars. Contours indicate 33, 67, 90, 
 and 99 per cent of each sample. The spatial extent and density of the entire RAVE-TGAS sample (i.e., all spectral types) is shown by the 2D histogram beneath. The white plus indicates the position of the Sun.}
 \label{fig:spatial_extent}
\end{figure}

We further restrict analysis to  stars that have:  a high signal-to-noise ratio
(\texttt{SNR\_K}); a reasonable estimate for their line-of-sight velocity
(\texttt{eHRV}); reliable atmospheric parameters because (i) the stellar-parameter
pipeline converged \citep[\texttt{Algo\_Conv\_K},][]{Kordopatis13}, and (ii) the
fit between the best-fitting model and the observed spectrum
from the chemical pipeline \citep[\texttt{CHISQ\_c},][]{Boeche11} was
reasonable. In addition, we remove peculiar stars flagged by the
classification pipeline as having anomalous spectra
\citep{Matijevic12}. Quantitatively, we require:
\begin{itemize}
	\item \texttt{SNR\_K} $> 40$
	\item \texttt{eHRV} $< 8\kms$
	\item \texttt{Algo\_Conv\_K} $\neq 1$
	\item \texttt{CHISQ\_c} $< 2000$
	\item \texttt{c1} = d, g, h, n, or o
	\item \texttt{c2} = d, g, h, n, o, or e
	\item \texttt{c3} = d, g, h, n, o, or e
\end{itemize}
We then apply the age 
bins described in Sec.~\ref{sec:comparison_age_bins}: young ($0
< \tau < 3\,$Gyr) and old ($8 < \tau < 13\,$Gyr), to 
this sample of turnoff stars. After applying these criteria, we are left with
6630 `young' stars and 5072 `old' stars. The spatial distributions of our selected young and old
populations are shown in Figure~\ref{fig:spatial_extent}, with blue and red
contours, respectively.

\subsubsection{Metallicity bins}
\label{sec:age_bins}

\begin{figure}
 \includegraphics[width=0.95\columnwidth]{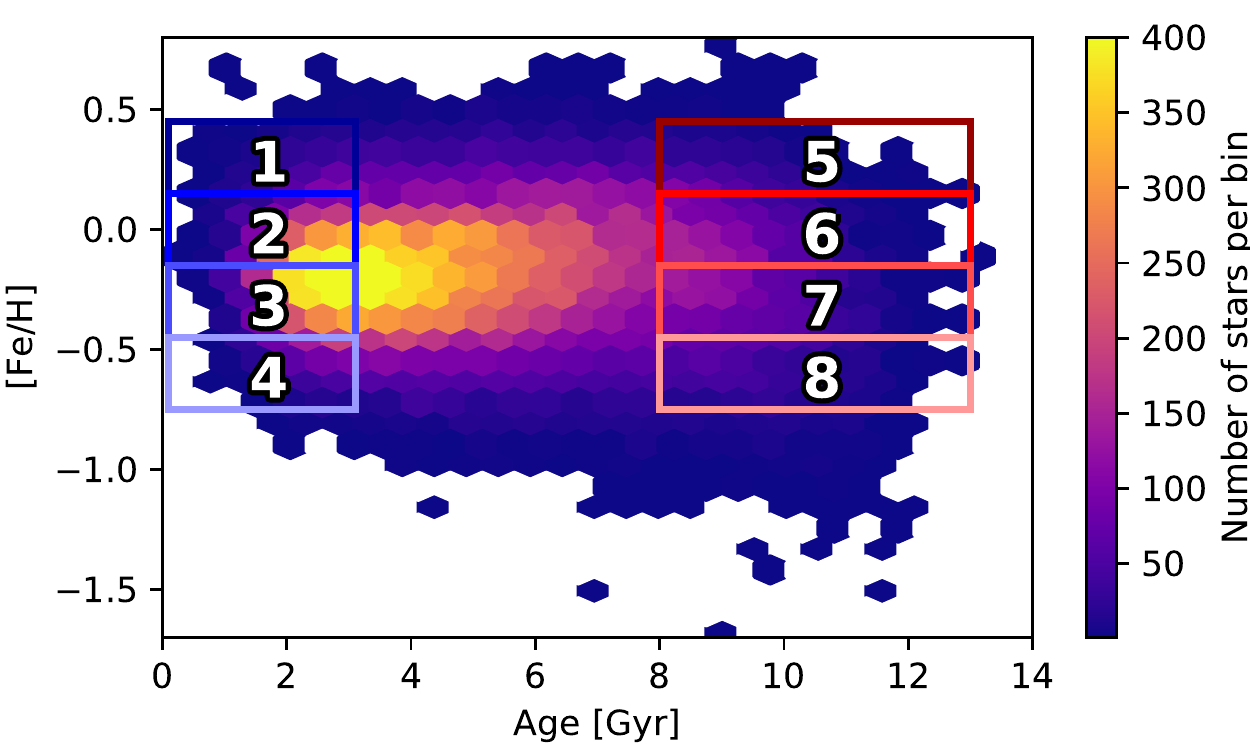}
 \caption{Age vs. [Fe/H] 2D histogram for our sample of turnoff stars defined by the red dashed lines in Figure~\ref{fig:teff_logg_age_err}, colour-coded by density. We consider four metallicity bins for each age group, described in Section~\ref{sec:age_bins}.}
 \label{fig:age_bins}
\end{figure}

We divide each age group into the four metallicity bins illustrated in
Figure~\ref{fig:age_bins}. The bins are:
\begin{itemize}
	\item Bins 1 and 5: $\enspace0.15 \leq \mathrm{[Fe/H]} < 0.45$
	\item Bins 2 and 6:$-0.15 \leq \mathrm{[Fe/H]} < 0.15$
	\item Bins 3 and 7:$-0.45 \leq \mathrm{[Fe/H]} < -0.15$
	\item Bins 4 and 8:$-0.75 \leq \mathrm{[Fe/H]} < -0.45$
\end{itemize}

\noindent 
Throughout the rest of the text, we will refer to these bins by
the numbers listed above and shown  in Figure~\ref{fig:age_bins}.
{We note that the radial distribution of stars does not vary significantly between these metallicity bins, for both the young 
(Bins 1-4) and old (Bins 5-8) samples. Our young sample has a Galactocentric radial distribution that peaks at $R \sim 
8.2$ kpc, with $\sigma_{R} \sim 0.25$ kpc, and for old stars, $R \sim 8.25$ kpc and $\sigma_{R} \sim 0.12$ kpc (see 
Figure~\ref{fig:spatial_extent}).}

\begin{figure*}
\centering
 \includegraphics[width=\textwidth]{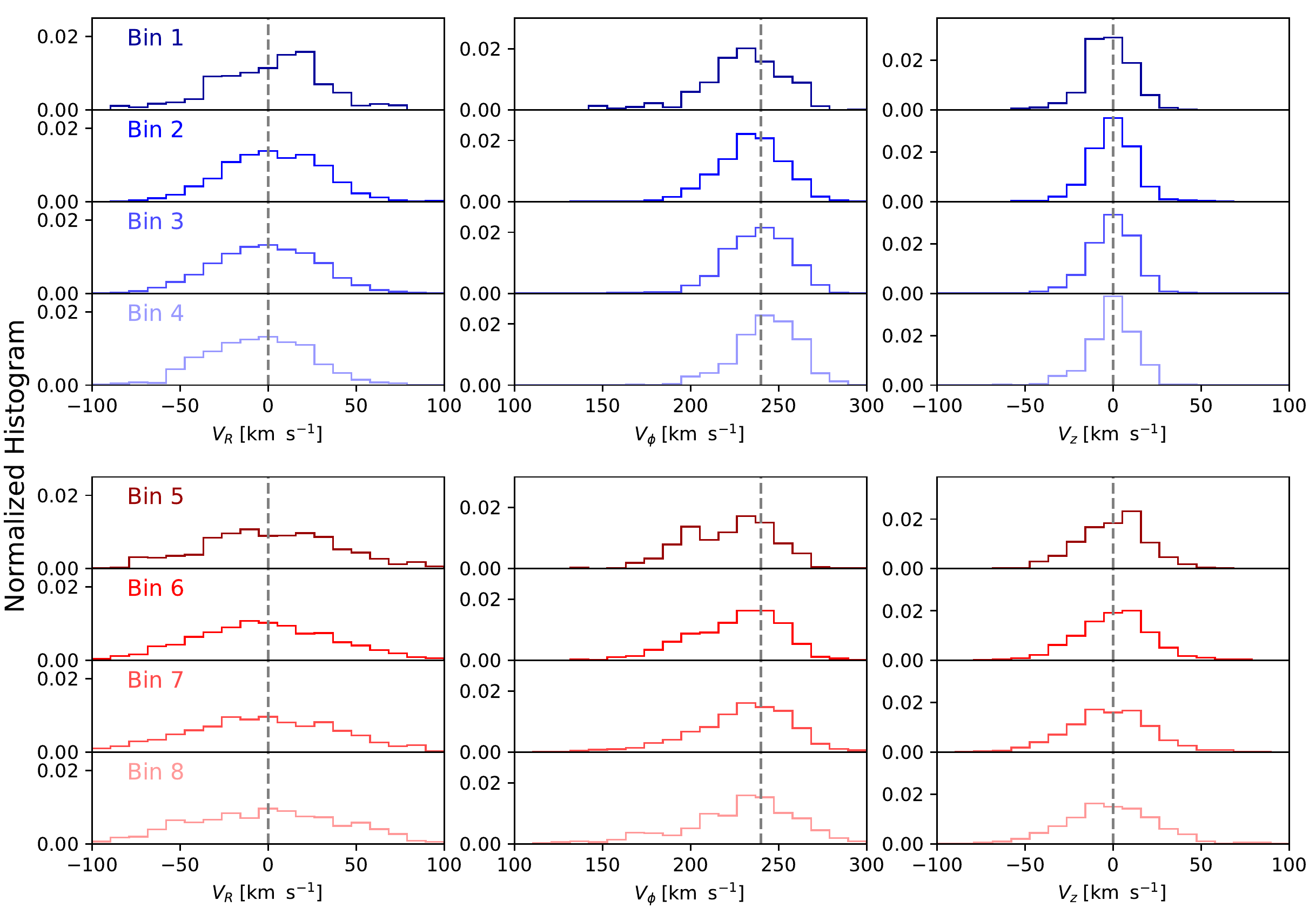}
 \caption{Cylindrical velocity distributions (left to right: $V_{R}, V_{\phi}, V_{z}$), weighted by the selection function, for our two age 
 bins. Young stars are plotted in blue, old stars are plotted in red. More metal-rich bins are plotted with 
 darker colours, with more metal-poor bins in lighter colours (metallicity Bins 1, 2, 3, and 4 for young stars, 
 and Bins 5, 6, 7, and 8 for old stars). Grey dashed lines indicate $0\kms$ for the $V_{R}$ and $V_{z}$ components, and $240\kms$ for the $V_{\phi}$ component.}
 \label{fig:cyl_vel_agebins}
\end{figure*}

\begin{table*}
\centering
\caption{Mean and dispersion for each component of the Galactocentric cylindrical velocity, weighted by the selection function, for both age groups and each metallicity bin within the age groups. }
\label{table:hist_values}
\begin{tabular}{llllllllll}
Bin number & Metallicity range                    & N    & $\overline{V_{R}}$     & $\sigma_{R}$  &$\overline{V_{\phi}}$      & $\sigma_{\phi}$  & $\overline{V_{z}}$      & $\sigma_{z}$  \\
\hline
Young & & & & & \\
1          & $0.15 \leq \mathrm{[Fe/H]} < 0.45$   & 386  & $0.5\pm1.3$                 & $30.3\pm1.1$                & $229.3\pm1.0$            & $23.6\pm0.8$               & $-2.4\pm0.5$             & $12.8\pm0.5$               \\
2          & $-0.15 \leq \mathrm{[Fe/H]} < 0.15$  & 2534 & $2.1\pm0.4$                & $27.2\pm0.4$                & $234.0\pm0.3$            & $18.4\pm0.3$                & $-0.2\pm0.2$               & $12.9\pm0.2$             \\
3          & $-0.45 \leq \mathrm{[Fe/H]} < -0.15$ & 3141 & $-1.8\pm0.4$               & $30.1\pm0.4$                & $237.7\pm0.3$            & $18.6\pm0.2$               & $-0.3\pm0.2$               & $13.2\pm0.2$             \\
4          & $-0.75 \leq \mathrm{[Fe/H]} < -0.45$ & 542  & $-5.6\pm1.0$               & $29.2\pm0.9$                & $243.2\pm0.6$            & $18.0\pm0.5$                & $0.1\pm0.5$               & $12.6\pm0.4$               \\
 & & & & & &\\
Old & & & & &\\
5          & $0.15 \leq \mathrm{[Fe/H]} < 0.45$   & 531   & $0.4\pm1.4$                 & $39.3\pm0.6$              & $222.1\pm0.9$             &  $24.4\pm0.9$               & $0.1\pm0.7$                & $19.0\pm1.2$             \\
6          & $-0.15 \leq \mathrm{[Fe/H]} < 0.15$  & 1757  & $-2.2\pm0.8$               & $40.5\pm1.0$                & $225.6\pm0.5$            & $25.8\pm1.6$                & $0.5\pm0.4$                & $21.2\pm2.0$               \\
7          & $-0.45 \leq \mathrm{[Fe/H]} < -0.15$ & 1740  & $-2.6\pm0.8$              & $43.2\pm1.0$                & $228.7\pm0.6$            & $28.9\pm1.4$               & $-2.2\pm0.5$               & $24.3\pm1.7$               \\
8          & $-0.75 \leq \mathrm{[Fe/H]} < -0.45$ & 838  & $-2.5\pm1.3$               & $47.2\pm0.6$                & $227.5\pm0.9$             & $32.9\pm0.9$                & $-1.6\pm0.8$                 & $27.3\pm1.0$               
\end{tabular}
\end{table*}

\section{Velocity distributions}
\label{sec:velocity_dist}

We now examine for each of the eight age--metallicity bins marked in
Fig.~\ref{fig:age_bins} the distributions of $V_R$, $V_\phi$ and $V_z$. We determine Galactocentric velocities in cylindrical coordinates following
Appendix A of \citet{Williams13} with the location of the Sun taken to be
$(R_0, z_0) = (8.3, 0)\,$kpc, and the local circular speed to be $V_{\rm LSR}
= 240\,$km s$^{-1}$ \citep{Schoenrich12}, and the solar peculiar velocity to be $(U, V, W)_{\odot}= (11.10, 12.24,
7.25)\,$km s$^{-1}$ \citep{Schoenrich10}.  $V_{R}$ is positive for motion
away from the Galactic centre, $V_{\phi}$ is positive in the direction of
Galactic rotation, and $V_z$ is positive towards the north Galactic pole.
Line-of-sight velocities are provided by RAVE DR5 \citep{Kunder17}, and
proper motions are taken from the TGAS catalogue \citep{Lindegren16}.

 \subsection{Correcting by the selection function of RAVE}
\label{sec:completeness}

As RAVE is not a volume-complete survey, when computing  kinematic trends
we need to correct for the selection function. \citet{Wojno17}
and references therein show that the selection function of RAVE is
relatively simple and well-behaved in the sense that RAVE is not kinematically
or chemically biased wherever its stellar parameter
pipeline gives reliable results. However, our sample comprises exclusively stars
contained in both RAVE and TGAS, which had vastly different observing
strategies and sky completeness, and therefore we had to reevaluate the selection
function $S_{\rm select}$. 
We computed $S_{\rm select}$ as a function of
position on the sky in \textsc{healpix} \citep{Gorski05} pixels,
$I_{\mathrm{2MASS}}$ magnitude, and ($J - K_s$) colour from the definition
\begin{equation}
\label{eq:completeness}
  S_{\rm select} (\rm pixel_{\alpha,\delta}) =
    \frac{\sum \sum N_{\rm RAVE}(\rm pixel_{\alpha,\delta},I,J-K_s)}{\sum \sum N_{\rm 2MASS}(\rm pixel_{\alpha,\delta},I,J-K_s)} .
\end{equation}
Each star contributes to the statistical analysis with a weight $w_i=1/
S_{\rm select}$ in the sense that the weighted mean $\overline{x}$ and weighted dispersion
$\sigma_x$ of an observable $x$ are
\begin{eqnarray}
\label{eq:weighted_mean}
  \overline{x} &=& \frac{\sum\limits_{i = 1}^{n} (x_i \times
  w_i)}{\sum\limits_{i = 1}^{n} w_i}\\
\label{eq:weighted_disp}
  \sigma_{x}^{2} &=& \frac{\sum\limits_{i = 1}^{n} w_i(x_i -
  \overline{x})^{2}}{k \sum\limits_{i = 1}^{n} w_i},
\end{eqnarray}
 where $k = (N' - 1) / N'$, and $N'$ is the number of non-zero weights. 

\subsection{Velocity trends with age and [Fe/H]}

Figure~\ref{fig:cyl_vel_agebins} shows (left to right) the weighted distributions of
$V_{R}, V_\phi$ and $V_z$ for stars in the young bin (blue, top
row), and the old bin (red, bottom row), with each age bin divided into the
four metallicity bins defined in Fig.~\ref{fig:age_bins}. Statistics for each
histogram, computed according to equations (\ref{eq:weighted_mean}) and
(\ref{eq:weighted_disp}), are given in Table~\ref{table:hist_values}.
We note that the values of  $\sigma_{R,\phi,z}$ given in Table~\ref{table:hist_values} have
been corrected for the contribution of observational uncertainties: 
\begin{equation}
\sigma_{R,\phi,z} = \sqrt{\sigma_{R,\phi,z}^{2} - \mathrm{e}_{V_{R,\phi,z}}^{2}} 
\end{equation} 
where $\mathrm{e}_{V_{R,\phi,z}}$ is the reported uncertainty for each
component of the velocity, weighted by the selection function.

 \begin{figure*}
\includegraphics[width=\textwidth]{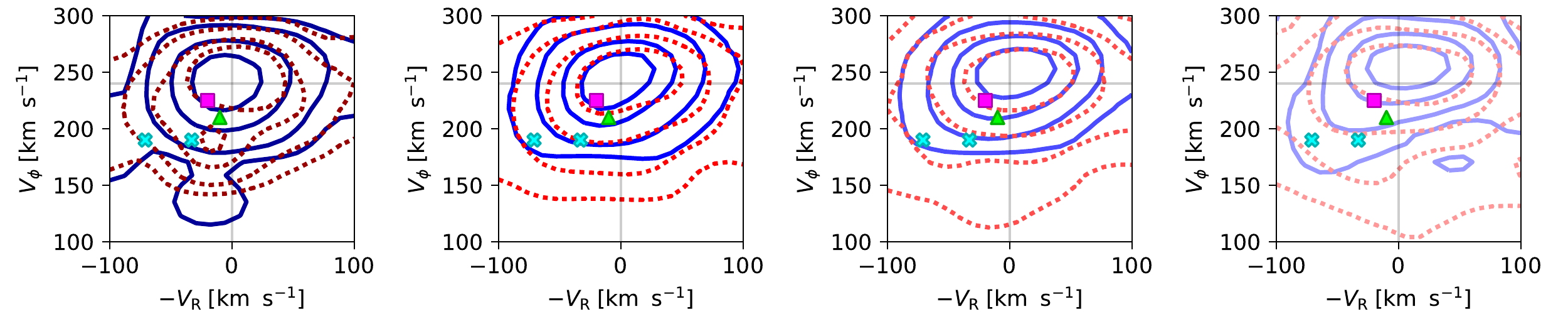}
 \caption{$-V_{R};V_{\phi}$ space for our sample of young (blue, solid) stars and old (red, dashed) stars, split into the four 
metallicity bins shown in Figure~\ref{fig:age_bins}. Contours show the density distributions for 33, 67, 90, and 99 per cent of 
the sample for each bin. The most metal-rich bin ($0.15 \leq \mathrm{[Fe/H]} < 0.45$) is on the left, and the most metal-poor bin 
($-0.75 \leq \mathrm{[Fe/H]} < -0.45$) is on the right. Cyan crosses indicate the two peaks in the kinematics of the Hercules moving group. The 
Hyades and Pleiades moving groups are indicated with a magenta square and a green triangle, respectively.}
 \label{fig:vr_vphi}
\end{figure*}

The values in Table~\ref{table:hist_values} for young stars (upper block)
show that ${V_{R}}$ and ${V_{z}}$ do not
vary significantly with metallicity. By contrast ${V_{\phi}}$
increases as [Fe/H] decreases. An inverse correlation of $V_\phi$ with [Fe/H]
within the thin disc is well known
\citep[e.g.][]{Lee11,Adibekyan13,Recio-Blanco14,Wojno16,Kordopatis17}, and
recognised to be a signature of the metallicity gradient within the thin disc
\citep[e.g.][]{Schoenrich10}: metal-poor stars tend to have large
guiding-centre radii and to visit us near pericentre, whereas metal-rich
stars tend to have small guiding-centre radii and visit us near apocentre. 
Quantitatively, we find
\begin{equation}
\frac{\partial{V_{\phi}}}{\partial\hbox{[Fe/H]}}\approx -15
\kms\,\hbox{dex}^{-1} \quad\hbox{(young stars)}.
\end{equation}
 {We estimate this value by a simple linear fit to the values of $\overline{V_{\phi}}$ for the four metallicity bins, taking the 
center of the metallicity bin for the value on the x-axis.}  Previous estimates of $\partial V_{\phi}/\partial$[Fe/H] have ranged from $-23 \kms
\mathrm{dex}^{-1}$ to $-11 \kms \mathrm{dex}^{-1}$. While our estimate is comparable, 
literature values usually indicate a steeper gradient than we find.

Table~\ref{table:hist_values} indicates that $\sigma_{\phi}$ decreases as
metallicity decreases, with Bin 1 having a dispersion $\sigma_{\phi} =
23.6\pm0.8 \kms$ and Bin 4 having $\sigma_{\phi} = 18.0\pm0.5 \kms$. This
trend is a natural corollary of an increase in mean Galactocentric radius
with decreasing metallicity and the stellar disc becoming cooler as one moves
outward. The data for $\sigma_R$ suggest, however, that we treat the trend in
$\sigma_\phi$ with caution because they do not show a corresponding decrease
while dynamical theory requires $\sigma_R$ and $\sigma_\phi$ to vary
together.

\begin{figure*}
 \includegraphics[width=\textwidth]{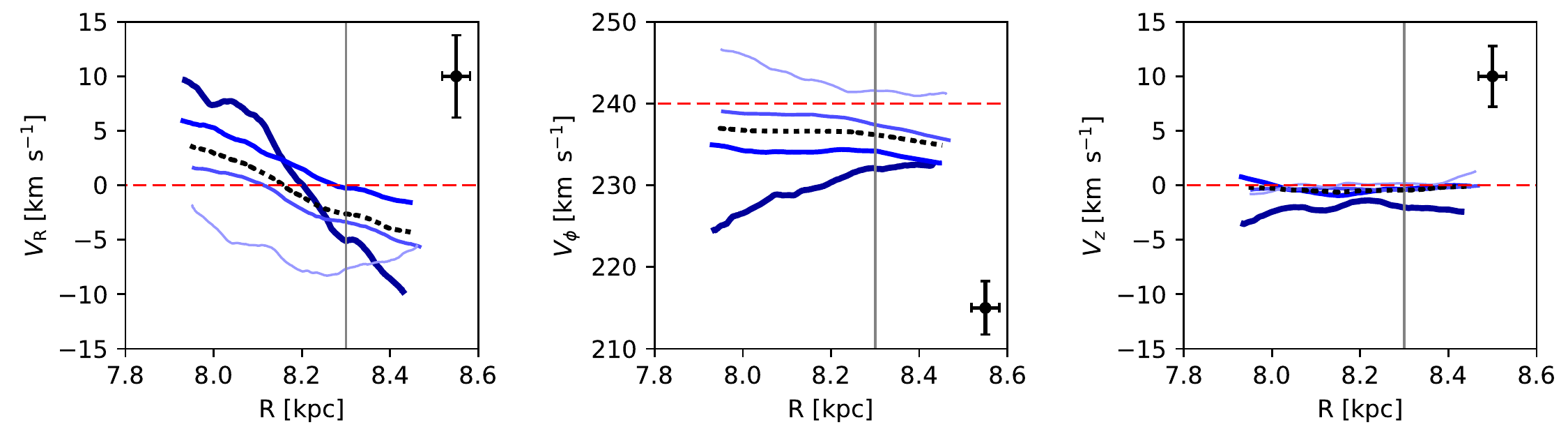}
\caption{Velocity trends as a function of Galactocentric radius (left to right: $V_{R}, V_{\phi}, V_{z}$), weighted by the 
selection function, for our young stars. The trend for the entire young sample ($-0.75 \leq \mathrm{[Fe/H]} < 0.45$) is indicated 
with the black dotted line. The most metal-rich bin ($0.15 \leq \mathrm{[Fe/H]} < 0.45$) is indicated with {the thick, dark blue line}, and the most 
metal-poor bin ($-0.75 \leq \mathrm{[Fe/H]} < -0.45$) with {the thin, light blue line}. The solid grey line indicates the position of the Sun. Average 
uncertainties are indicated in each panel by the error bar.}
 \label{fig:age_vel_R_young}
\end{figure*}

In line with much previous work \citep[e.g.,][]{Nordstroem04,Guiglion15} the histograms
for old stars are broader than those for young stars. All three velocity
dispersions increase with decreasing [Fe/H], so amongst old stars the
metal-poor component is the hottest -- this is the reverse of the trend in
$\sigma_\phi$ that we just saw in the young stars. Gratifyingly, the data for
${V}_\phi$ for old stars show it to decrease as [Fe/H] increases,
just as dynamical theory requires given the increase in $\sigma_\phi$.
Many previous studies \citep[e.g.][]{Spagna10,Kordopatis11, Adibekyan13,
Kordopatis13_b,Wojno16,Kordopatis17} have found the same trend of
${V}_\phi$ with [Fe/H] but they have generally obtained a steeper
gradient than our value
\begin{equation}
\frac{\partial V_{\phi}}{\partial\hbox{[Fe/H]}} \approx 5 \kms\,
\mathrm{dex}^{-1}\quad\hbox{(old stars)}.
\end{equation}
Measurements from the literature range from $42
\kms\,\mathrm{dex}^{-1}$ to $51 \kms\,\mathrm{dex}^{-1}$.
The literature values are much larger than ours, probably because they refer
to samples of thick disc stars (selected chemically, kinematically, or spatially), 
whereas we have imposed no such selection. 

Figure~\ref{fig:vr_vphi} shows the distribution of young (blue) and old (red)
stars in the $(-V_R,V_\phi)$ plane, which is the natural extension of the
traditional $(U,V)$ plane. The histograms in the left and central columns of
Fig.~\ref{fig:cyl_vel_agebins} are projections of these distributions onto
the $V_R$ and $V_\phi$ axes. In Fig.~\ref{fig:vr_vphi} metallicity increases
from left to right.
Literature values for local maxima associated with the Hercules (cyan
cross), Hyades (magenta square), and Pleiades (green triangle) moving groups
are overplotted \citep{Dehnen00,BT08}. 

The peak in the distribution of young stars moves up and to the right with
decreasing metallicity, as we expect given that for these stars $\partial
V_\phi/\partial\hbox{[Fe/H]}<0$ and $\partial V_R/\partial\hbox{[Fe/H]}>0$.  With the
possible exception of the Hercules stream (cyan cross) the contours do not
provide convincing evidence for the streams detected by \citet{Dehnen00} in
Hipparcos data. Curiously, the signature of the Hyades stream is most
convincing for the most metal-poor Bins 4 and 8. Given the small size of the
sample of old stars, and the way the red contours change 
from one metallicity panel to another, we conclude that these
wiggles probably owe more to noise than moving groups.
 
The histograms for $V_z$ in the right column of
Fig.~\ref{fig:cyl_vel_agebins} show that, as expected, the old stars are
vertically hotter than the young stars. Whereas for the old stars $\sigma_z$ tends to increase
with decreasing metallicity, from $\sigma_{z} =
19.0\pm1.2 \kms$ for Bin 5 to $\sigma_{z} =
27.3\pm1.0 \kms$ for Bin 8, $\sigma_z$ is essentially
independent of metallicity for the young stars. The increase for old stars in $\sigma_z$ with
decreasing [Fe/H], like the decrease in $V_\phi$ with decreasing [Fe/H], points
to old, metal-poor stars being on highly eccentric orbits that are not
tightly confined to the plane. 

\subsection{Velocity trends with $R$}
\label{sec:velocity_trends}

\begin{figure*}
 \includegraphics[width=\textwidth]{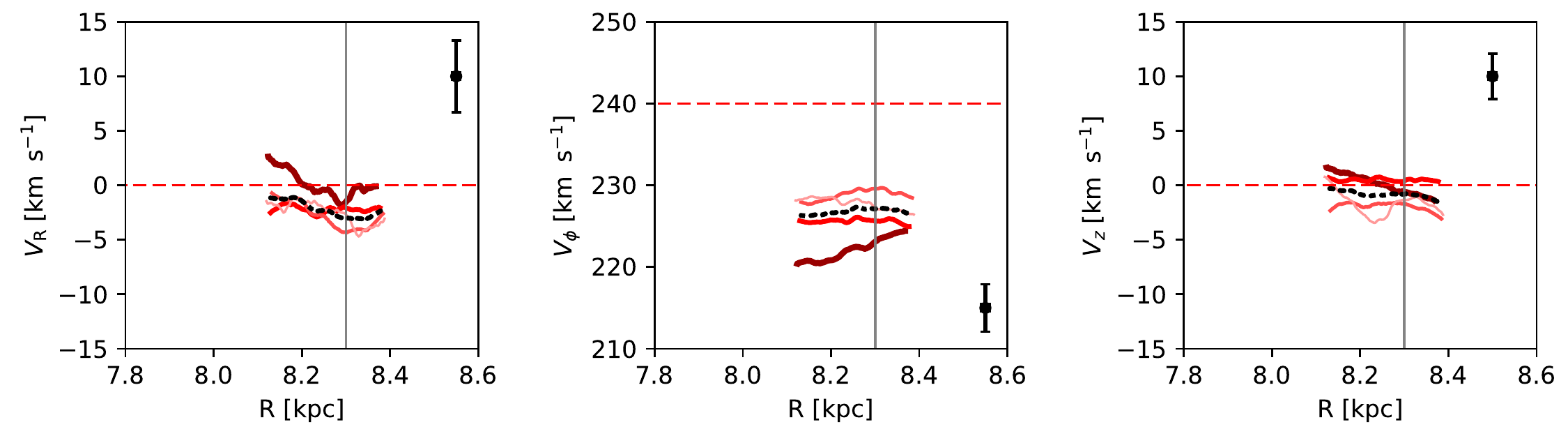}
 \caption{Velocity trends as a function of Galactocentric radius (left to right: $V_{R}, V_{\phi}, V_{z}$), weighted by the 
 selection function, for our old stars. The trend for the entire old sample ($-0.75 \leq \mathrm{[Fe/H]} < 0.45$) is indicated with the 
 black dotted line. The most metal-rich bin ($0.15 \leq \mathrm{[Fe/H]} < 0.45$) is indicated {with the thick, dark red line}, and the most metal-
poor bin ($-0.75 \leq \mathrm{[Fe/H]} < -0.45$) with {the thin, light pink line}. The solid grey line indicates the position of the Sun. Average 
 uncertainties are indicated in each panel by the error bar.}
 \label{fig:age_vel_R_old}
\end{figure*}

We now consider kinematic trends with Galactocentric radius.  In
Figures~\ref{fig:age_vel_R_young} and \ref{fig:age_vel_R_old}, we show
$V_{R}, V_{\phi}, V_{z}$ as functions of $R$ for the young and old
stars, respectively. Metal-rich bins are shown in darker colours {with thicker lines},
with successively lighter colours {and thinner lines} indicating decreasing metallicity. The plotted values are calculated as follows. The stars
in a given age-metallicity bin were ordered by $R$ and then a window
containing $N$ stars was moved along the resulting lineup, and the mean of
$V_R$, etc., was computed at each location of the window. This operation was
carried out for 1000 values of $N$ that were drawn from a uniform distribution with a 
minimum of 20, and a maximum which varied as a function of 
the size of the age-metallicity bin. The results were then averaged.  Average uncertainties for each velocity component are
given in their respective panels. The uncertainty in $R$ largely correlates
with the range in Galactocentric radius sampled, and is of the order of
$\sim0.03$ and $\sim0.01$ kpc for young and old stars, respectively. 

To check that this procedure generates velocity gradients that are consistent
with published gradients, we used it to compute velocity gradients for (i)
the complete set of turnoff stars, i.e. all stars in the red dashed rectangle
of Figure~\ref{fig:teff_logg_age_err}, regardless of age or metallicity, and
(ii) the set of red giant stars ($2<\log g<2.5$) that satisfy the quality
criteria given in Section~\ref{sec:RAVE_qc}.  For the turnoff stars we find
$\partial V_R/\partial R=-5.7\pm2.0\kms\,\hbox{kpc}^{-1}$, while for the
giant stars $\partial V_R /\partial R = -5.4\pm2.1\kms\,\hbox{kpc}^{-1}$.
Both of these values agree well with literature values:
$-3\kms\,\hbox{kpc}^{-1}$ from \citet{Siebert11}, $-8\kms\,\hbox{kpc}^{-1}$
from \citet{Williams13}, and $-6.6\pm0.7\kms\,\hbox{kpc}^{-1}$ from
\citet{Bovy17}. Negative values of $\partial V_R/\partial R$ signal that
we lie in a region in which the stellar fluid is being compressed as
stars at both smaller and larger radii are moving towards us.

\subsubsection{Young stars (\,$0 < \tau < 3$ Gyr)}

The dashed black lines in Figure~\ref{fig:age_vel_R_young} show the velocity
trends for our entire sample of young stars, summed over all four metallicity
bins.  The gradient of the curve in the panel for $V_R$ is $\partial
V_R/\partial R = -11.4\pm4.0\kms\,\hbox{kpc}$.  When the sample is decomposed
by metallicity, the gradient $\partial V_R/\partial R$ of the most metal-rich
bin (1) is markedly steeper ($-28.0\pm9.9\kms\,\hbox{kpc}^{-1}$) than those
of Bins 2 and 3. The curve $V_R(R)$ for the most metal-poor bin (4) shifts
from a negative gradient at smaller radii to a positive gradient at the
largest values of $R$.  The gradients we obtain for Bins 1, 2, and 3 are all
steeper than typical literature values for more inhomogeneous populations
($\lesssim-3\kms\,\hbox{kpc}$ from \citet{Siebert11};
$-8\kms\,\hbox{kpc}^{-1}$ from \citet{Williams13};
$-6.6\pm0.7\kms\,\hbox{kpc}^{-1}$ from \citet{Bovy17}; $-7.01\pm
0.61\kms\,\hbox{kpc}^{-1}$, and $-9.42 \pm
1.77\kms\,\hbox{kpc}^{-1}$ from \citet{Carrillo18} for measurements made
below and above the plane of the disc, respectively). 

The central panel of Fig.~\ref{fig:age_vel_R_young} displays again the
familiar result that at any radius $V_\phi$ decreases with increasing [Fe/H]
(upper central panel of Fig.~\ref{fig:cyl_vel_agebins}). The new result we
take from Fig.~\ref{fig:age_vel_R_young} is that the sign of $\partial
V_\phi/\partial R$ changes from positive for the metal-rich Bin 1 to negative
for the metal-poor Bin 4. This trend among our young stars is consistent with
a finding of \citet[][their Figure 8]{Rojas-Arriagada16} for the thin disc as
a whole.

\begin{table}
\centering
\caption{Measured radial velocity gradients for each age group and metallicity bin.}
\label{my-label}
\begin{tabular}{lll}
Bin Number & Metallicity range & $\partial {V_{{R}}} /\partial R$  \\
  & &  \;km\,s$^{-1}$\,kpc$^{-1}$ \\
\hline
Young      &                                                         \\
1          &$0.15 \leq \mathrm{[Fe/H]} < 0.45$& $-28.0\pm9.9$                 \\
2          &$-0.15 \leq \mathrm{[Fe/H]} < 0.15$& $-10.6\pm3.8$                 \\
3          &$-0.45 \leq \mathrm{[Fe/H]} < -0.15$& $-10.2\pm3.6$                 \\
4          &$-0.75 \leq \mathrm{[Fe/H]} < -0.45$& $-3.6\pm1.3$                 \\
all	   & $-0.75 \leq \mathrm{[Fe/H]} < 0.45$& $-11.4\pm4.0$		\\
           &                                                         \\
Old        &                                                         \\
5          &$0.15 \leq \mathrm{[Fe/H]} < 0.45$& $-5.9\pm2.5$                 \\
6          &$-0.15 \leq \mathrm{[Fe/H]} < 0.15$& $-0.7\pm1.1$                  \\
7          &$-0.45 \leq \mathrm{[Fe/H]} < -0.15$& $-9.9\pm4.2$                  \\
8          &$-0.75 \leq \mathrm{[Fe/H]} < -0.45$& $-6.8\pm2.8$                 \\
all	   &$-0.75 \leq \mathrm{[Fe/H]} < 0.45$& $-5.9\pm 2.6$		\\
\end{tabular}
\end{table}

The right panel of Fig.~\ref{fig:age_vel_R_young} shows that $V_z$ is
consistent with zero at all radii for all metallicity bins.

\subsubsection{Old stars (\,$8 < \tau < 13$ Gyr)}

Figure~\ref{fig:age_vel_R_old} shows the dependence of $\mathbf{V}$ on $R$
for old stars. Unfortunately, the old stars probe a significantly smaller
range in $R$ than do the young stars. 

When all four metallicity bins are aggregated, we obtain a
negative gradient, $\partial {V_{R}}/\partial R =
-5.9\pm2.6\kms\,\hbox{kpc}^{-1}$. There is a suggestion that the mean value
of $V_R$ decreases with decreasing metallicity, but there is no evident
variation of $\partial V_R/\partial R$ with metallicity, contrary 
to what we found for the young stars.

All four metallicity bins of old stars have lagging values of
$V_\phi\sim-14\kms$, as
we expect given the relatively large velocity dispersions of these
populations. Any dependence on metallicity of the gradient $\partial
V_\phi/\partial R$ is, for the most part, lost in the noise, contrary to what we found for
the young stars. However, we do find that the most metal-rich bin (5) has the greatest
lag, similar to Bin 1 for young stars.

The old stars, like the young stars, show no systematic dependence of $V_z$
on either metallicity or radius, and for all populations $V_z$ is consistent
with zero.

\section{Orbital parameters}
\label{sec:orbital_parameters}
An orbit in the MWPotential2014 potential was computed for each star using
the \textsc{galpy} \citep{Bovy15_galpy} python package. From the orbit we
tabulated the eccentricity $e$, guiding radius ($R_G$), and the maximum height
above the plane $z_{\rm max}$. 

\begin{figure}
 \includegraphics[width=0.8\columnwidth]{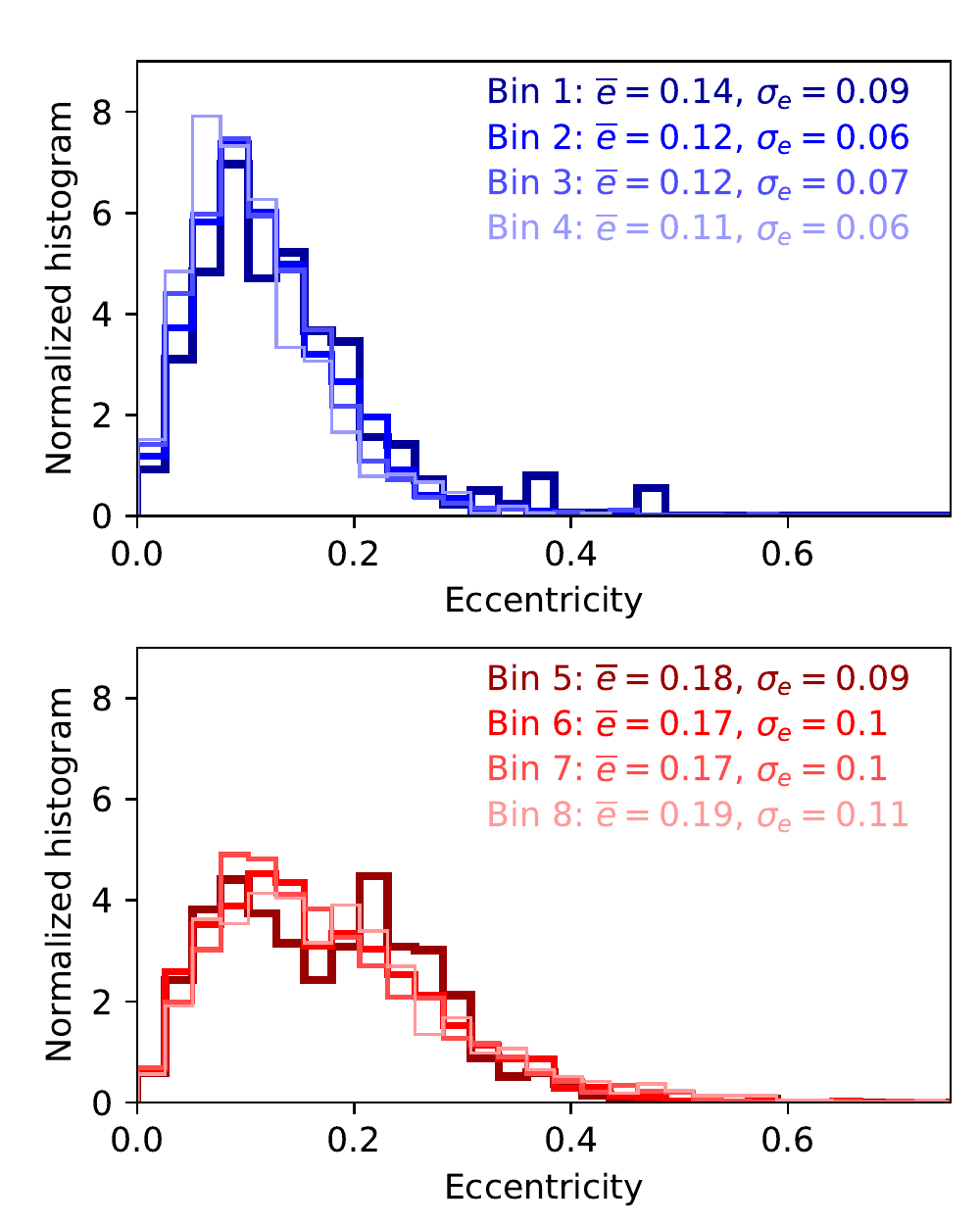}
 \caption{Orbital eccentricities for our sample of young (top) and old (bottom) stars. Bins in metallicity are defined in Section~\ref{sec:age_bins}.}
 \label{fig:age_ecc}
\end{figure}

\begin{figure}
 \includegraphics[width=0.8\columnwidth]{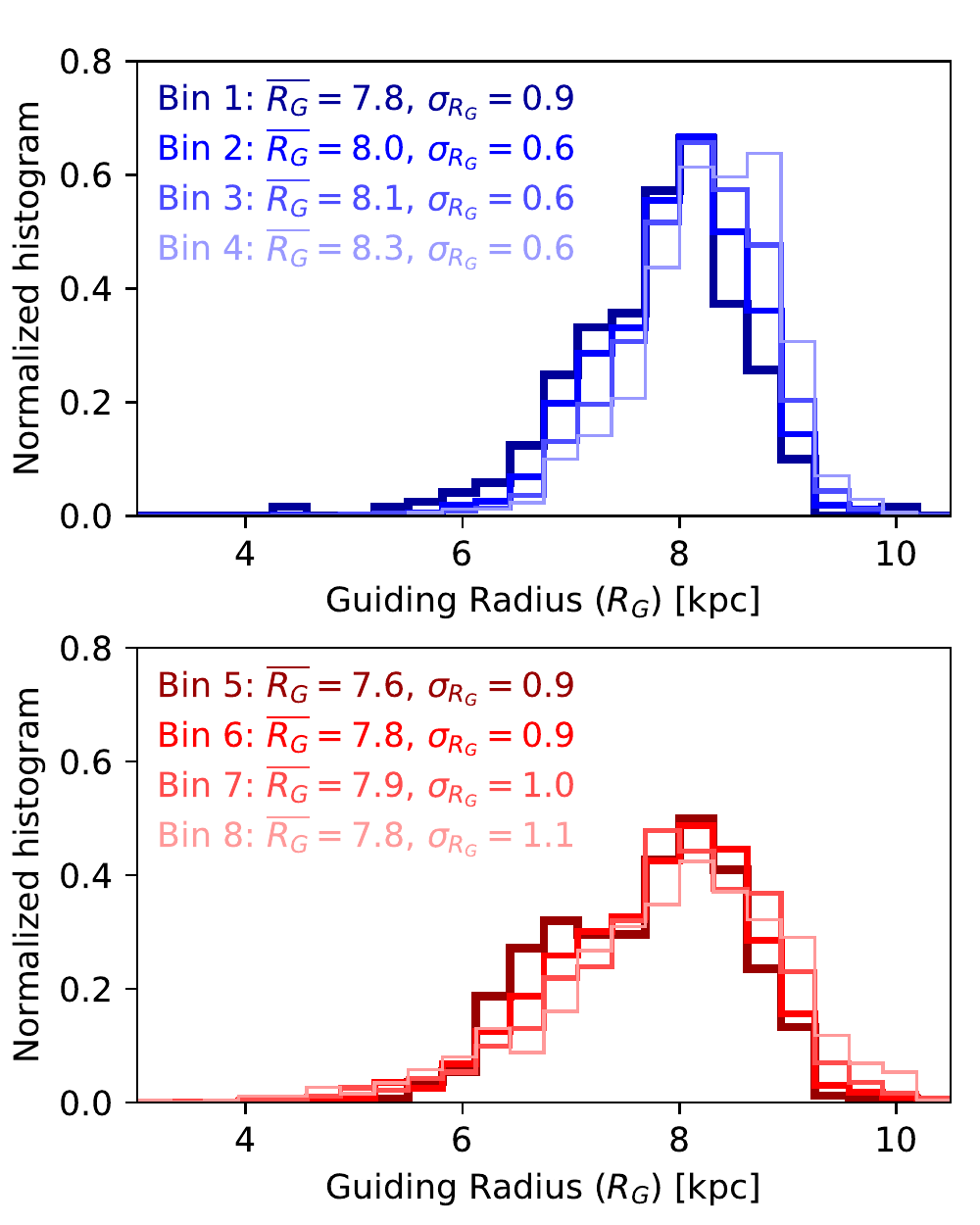}
 \caption{Maximum height above the plane ($Z_{\rm max}$) for our sample of young (top) and old (bottom) stars. Bins in metallicity are defined in Section~\ref{sec:age_bins}.}
 \label{fig:age_rg}
\end{figure}

\begin{figure}
 \includegraphics[width=0.8\columnwidth]{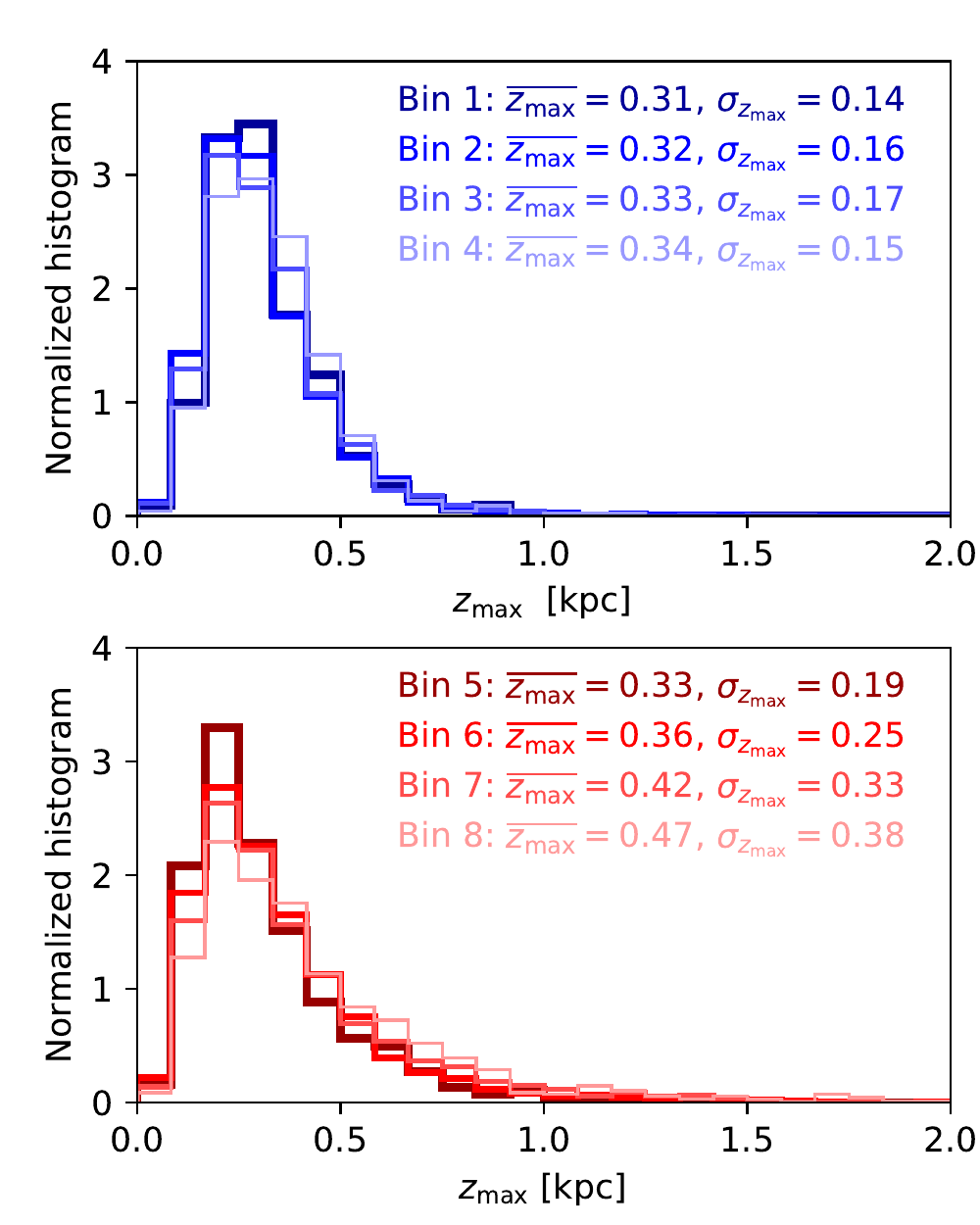}
 \caption{Maximum height above the plane ($Z_{\rm max}$) for our sample of young (top) and old (bottom) stars. Bins in metallicity are defined in Section~\ref{sec:age_bins}.}
 \label{fig:age_zmax}
\end{figure}

\subsubsection{Eccentricities}

In Figure~\ref{fig:age_ecc}, we show the
eccentricity distributions for our young (left panel) and old stars (right
panel), divided into bins of metallicity and weighted by the factors $w_i$
defined in Section~\ref{sec:completeness}. When all four metallicity bins are aggregated, the eccentricity distribution
of our young stars peaks at $\overline{e} = 0.12$ and has dispersion $\sigma_{e} =
0.07$ (typical of the thin disc).  The mean value of $e$
shifts to smaller values with decreasing metallicity ($\overline{e} =
0.14 $ for Bin 1 and $\overline{e} = 0.11$ for Bin 4), and the dispersion
of $e$ decreases slightly with decreasing metallicity. 

The old stars have a much broader
distribution in $e$, and consequently a larger mean eccentricity. 
When the metallicity bins are aggregated, the mean eccentricity of old stars
is $\overline{e} = 0.18$ and its dispersion is $\sigma_{e} = 0.1$
(typical of the thick disc, see \citealt{Kordopatis11}). In contrast to the
young population, the mean and dispersion of the eccentricity distributions are very similar,
and slightly increase with decreasing metallicity: the mean $e$ increases from
$\overline{e} = 0.18$ for Bin 5 to $\overline{e} = 0.19$ for Bin 8. {We note that for the most metal-rich bin 
($0.15 \leq \mathrm{[Fe/H]} < 0.45$), the histogram is suggestive of a bimodality in the eccentricity distribution.}
Uncertainties in $\overline{e}$ and $\sigma_{e}$ are of the order
$\sim 0.01$.

\subsubsection{Guiding radii}
The distributions of guiding radii for our young and old samples are shown in 
Figure~\ref{fig:age_rg}. Young stars have mean guiding radii closer to the solar neighbourhood ($\overline{R_G} = 7.8-8.3$ kpc) compared to old 
stars, which have guiding radii towards the inner disc ($\overline{R_G} = 7.6-7.9$ kpc). The dispersion in guiding radii is also larger for old stars. {We also note a hint of a bimodality in the guiding radii distribution for old, metal-rich stars, similar to their  eccentricity distribution.} In addition, for young stars, 
we find an increase in guiding radii as a function of decreasing metallicity, corresponding to the results shown in  
Figure~\ref{fig:age_ecc}. Uncertainties in $\overline{R_G}$ and $\sigma_{R_G}$ are of the order
$\sim 0.03$ kpc.

\subsubsection{Maximum height above the plane}

In Figure~\ref{fig:age_zmax}, we show the distributions of $z_{\rm max}$
for our young and old stars.  All metallicity bins of young stars have similar means
($\overline{z_{\mathrm{max}}} = 0.31-0.34\,$kpc). For young stars,
uncertainties on $\overline{z_{\rm max}}$ and $\sigma_{z_{\rm
max}}$ are of the order $\sim 0.01\,$kpc. By contrast
the old stars show a signficant difference in mean $Z_{\rm max}$ between the most
metal-rich (Bin 5, $\overline{z_{\mathrm{max}}} = 0.3\,$kpc), and the most
metal-poor bin ($\overline{z_{\mathrm{max}}} = 0.52\,$kpc). Uncertainties
on $\overline{{z}_{\rm max}}$ and $\sigma_{z_{\rm max}}$ for old
stars are of the order $\sim 0.02\,$kpc.

The distributions of $z_{\rm max}$ for young and old stars are remarkably
similar. Both are very skew, but the distribution for old stars has a much
stronger tail towards high $z_{\rm max}$. In interpreting these distributions
of $z_{\rm max}$ it must be borne in mind that RAVE's sightlines avoid the
plane, so stars tend to be observed at significant values of $z$. Clearly
$z_{\rm max}\ge z$. Given the strong correlation of $\sigma_z$ with age
established in earlier work \citep[e.g.][]{Casagrande11}, most of the
youngest stars in the disc have no chance of entering the RAVE catalogue
because they do not move far enough from the plane. That is, RAVE must be
catching only the high-$V_z$ tail of the youngest stars.

\section{Discussion and Conclusions}
\label{sec:discussion}

The release of Gaia-based parallaxes for the majority of RAVE stars has
significantly enhanced the value of the RAVE catalogue by reducing errors in
distances and proper motions, particularly for main-sequence and turnoff
stars. The new parallaxes also make it possible to determine credible ages
for stars in the turnoff region.  It is well established that the kinematics
of a stellar population is correlated with age and chemistry. Since these
correlations arise from the secular evolution of the Galaxy's mass
distribution, star-formation rate and non-axisymmetric features
\citep[e.g.][]{Aumer16,Schoenrich17}, they must encode valuable
information about our Galaxy's history. Hence it is interesting to re-examine
the kinematics-age-metallicity nexus for turnoff stars that appear in both
the RAVE and TGAS catalogs.

Unfortunately, even for stars in the RAVE-TGAS overlap, ages are quite
uncertain, so we have restricted the analysis to the youngest ($\tau<3\,$Gyr)
and oldest ($\tau>8\,$Gyr) stars in the expectation that the given
uncertainties cannot scatter stars from one extreme group to the other.
A corollary of this restriction is that the studied samples are
rather small: 6630 young stars and 5072 old stars. We note that while we expect our young stars to be  
relatively free of contamination, the same cannot be said for the old stars, where we estimate $\sim 27$ per cent 
have true ages less than 6 Gyr. 

A feature of samples drawn from RAVE that one must always bear in mind is
that few RAVE stars lie close to the plane, and distance from the Sun and
distance from the plane are correlated. These features arise because
RAVE avoids sight lines at low Galactic latitudes. 
A significant corollary for our sample of young stars is that they must all be
outliers in the true distribution of vertical energy and they may not be
typical of young stars as a whole. 

While the young stars cover a range $\sim0.5\,$kpc in $R$, the sample of old,
less luminous stars, has not much more than half the radial range.
Consequently, it is hard to establish radial trends for the old stars.

{We have used hot turnoff stars to conduct our analysis, and here we note a small caveat regarding RAVE turnoff stars and their derived distances. \citet{Kunder17} found that distances to RAVE's hot dwarfs in DR5 were
some of the most problematic, and an erroneous distance can result in motion
associated with differential rotation artificially enhancing values derived
for $V_R$ \citep{SchoenrichBinneyAsplund12}. However, \citet{McMillan17}, using TGAS parallaxes as a prior 
to provide updated distance measurements, found that systematic uncertainties were greatly reduced for turnoff stars (see their 
Section 7). Therefore, while we cannot be completely sure that these results are not artifacts
induced by systematic errors in the distances to these hot turnoff stars, we have little reason to believe that our results are simply 
due to systematics.}

Overall, we find negative radial velocity gradients as a function of Galactocentric radii, and measure steeper gradients than 
those found before 
for RAVE stars \citep{Siebert11, Williams13}. The source of the negative gradient in $\partial {V_{\rm R}}/\partial R$ is typically 
attributed to flows induced by non-axisymmetries in the disc \citep[i.e., the bar and/or spiral arms,][]{Siebert11,Siebert12}. 
When we split each age group into bins of metallicity, we show, for the first time, that $\partial{V_{R}}/\partial R$ 
steepens with metallicity. While a signature of this trend is found in both the young and old stars, it is much 
more pronounced for young stars.

{A possible physical interpretation for this difference may be found by revisiting the $-V_R; V_{\phi}$ plane (Figure~\ref{fig:vr_vphi}, analogous to the 
$U;V$ plane). We recall that moving groups in the local velocity field have been associated with the effects of resonant gravitational 
interactions due to non-axisymmetries in the MW disc, e.g. the Hercules moving group is typically explained as a signature of a bar 
that perturbs the orbits of stars near its OLR \citep[e.g.][]{Dehnen00, Fux01}, and the Hyades moving group is 
typically associated with spiral structure \citep{Quillen05}. Figure~\ref{fig:vr_vphi} shows that for our sample, 
stars with similar kinematics to those of the Hyades and Hercules moving groups are seen in all metallicity bins and ages. 
However, we find that the central peak of the distribution for young stars shifts towards larger values of $-V_R; V_{\phi}$ as 
metallicity decreases. This shift in 
the peak of the kinematics of young stars may indicate that the relative contribution of each non-axisymmetric component of the 
disc (the spiral arms and bar, respectively) to the perturbed stellar kinematics varies as a function of metallicity. We note that this 
correlation also corresponds, as expected, to the observed metallicity gradient in the Galaxy (of the order $\partial$[Fe/H]/$\partial R \sim -0.06$ dex $\,\hbox{kpc}^{-1}$, \citealt{Boeche13b,Genovali14}).}

{The presence of young, metal-rich stars with similar kinematics as the Hyades moving group is consistent 
with the results of \citet{Quillen05,Famaey08,Antoja17}, who found the moving group to be metal-rich and produced through 
gravitational interactions with spiral arms. In addition, the fact that we see these metal-rich stars visiting the solar neighborhood from the inner disc, with similar kinematics as 
the Hercules moving group, suggests that these stars may have had their orbits affected by similar dynamical interactions with  
the central bar \citep{Ramya16,Antoja17}. Taking these points into account, we suggest that dynamical effects due to both the bar 
and spiral arms contribute to the steeper radial velocity gradient we find for young, metal-rich stars, while the more metal-poor stars 
are less affected by the bar.
Our findings are then roughly consistent with \citet[][see their Figure 4]{Monari17_df}, where moving further from the OLR decreases the contribution of the bar, similar to what we find for our more metal-poor stars.}

We also find secure results regarding the variation of $V_\phi$ with age,
metallicity and Galactocentric radius. All the old populations show a
significant lag of $V_\phi$ relative to the circular speed, with little dependence 
on metallicity. Among the 
young stars, it is the most metal-rich that show the greatest lag in $V_\phi$,
which is a manifestation of the metallicity gradient in the thin disc and epicyclic motions of stars. The most metal-rich subsample of young stars shows rather surprising trends
with $R$ in both $V_R$ and $V_\phi$: its value for $\partial V_R/\partial R$
is much more negative than the values we obtain for the less metal-rich parts
of the young sample, and its value for $\partial V_\phi/\partial R$ is
strongly positive while the more metal-weak young stars show weakly negative
values. 

We find that the super-solar metallicity young stars have, on average, lower eccentricities than the super-solar metallicity 
 old stars, pointing towards the possibility that the latter have migrated from much further in the Galactic disc. {This result 
 adds another dimension (age) to the findings of \citet{Kordopatis15_rich}, where they show that super metal-rich (SMR) stars in the 
 solar neighborhood are on relatively circular orbits. They conclude that these SMR stars must be predominantly affected by 
 resonant scattering at the OLR of the spiral arms \citep[churning,][]{Sellwood02}, as their orbital angular momentum has increased to bring them to the solar neighbourhood, without a corresponding increase in eccentricity. We also note the 
 bimodality in the eccentricity and guiding radii distributions for old, metal-rich stars, and suggest that this may be a possible signature of the different processes which bring them to the solar radius-- either through epicyclic motions (i.e., stars with 
 larger guiding radii which are temporarily visiting the solar neighbourhood), or churning (true migrators from the inner Galaxy).}
 
 {With Gaia DR2\footnote{https://www.cosmos.esa.int/web/gaia/dr2} and additional individual chemical abundances (Wyse et al., in 
 prep),} precise age estimates of these stars should allow for observational constraints on the relative efficiency of processes which 
 bring them to the solar neighbourhood. 
 
\section*{Acknowledgements}
We thank the anonymous referee for their suggestions that greatly improved the quality of the manuscript. We also thank Alice 
Quillen, Friedrich Anders, and Kyle Oman for their comments and enlightening discussions.
Funding for this work and for RAVE has been provided by: the
Australian Astronomical Observatory; the Leibniz-Institut
f\"ur Astrophysik Potsdam (AIP); the Australian National
University; the Australian Research Council; the European Research Council
under the European Union's Seventh Framework
Programme (Grant Agreement 240271 and 321067); the French National
Research Agency; the German Research Foundation
(SPP 1177 and SFB 881); the Istituto Nazionale di
Astrofisica at Padova; The Johns Hopkins University; the
National Science Foundation of the USA (AST-0908326);
the W. M. Keck foundation; the Macquarie University;
the Netherlands Research School for Astronomy; the Natural
Sciences and Engineering Research Council of Canada;
the Slovenian Research Agency; the Swiss National Science
Foundation; the Science \& Technology Facilities Council of
the UK; Opticon; Strasbourg Observatory; and the Universities
of Groningen, Heidelberg and Sydney. The RAVE web
site is at https://www.rave-survey.org. This work has made use of data from the European Space Agency (ESA)
mission {\it Gaia} (\url{https://www.cosmos.esa.int/gaia}), processed by
the {\it Gaia} Data Processing and Analysis Consortium (DPAC,
\url{https://www.cosmos.esa.int/web/gaia/dpac/consortium}). Funding for
the DPAC has been provided by national institutions, in particular the
institutions participating in the {\it Gaia} Multilateral Agreement.




\bibliographystyle{mnras}
\bibliography{mybib} 







\bsp	
\label{lastpage}
\end{document}